# Physics-Aware POD-Based Learning for *Ab initio* QEM-Galerkin Simulations of Periodic Nanostructures

Martin Veresko, Yu Liu, Daqing Hou[a], Ming-Cheng Cheng*

Department of Electrical and Computer Engineering
Clarkson University, Potsdam, NY 13699-5720

[a]Current Affiliation: Department of Software Engineering
Rochester Institute of Technology, Rochester NY 14623

*Corresponding Author: mcheng@clarkson.edu

## Abstract

Quantum nanostructures offer crucial applications in electronics, photonics, materials, drugs, etc. For accurate design and analysis of nanostructures and materials, simulations of the Schrödinger or Schrödinger-like equation are always needed. For large nanostructures, these eigenvalue problems can be computationally intensive. One effective solution is a learning method via Proper Orthogonal Decomposition (POD), together with *ab initio* Galerkin projection of the Schrödinger equation. POD-Galerkin projects the problem onto a reduced-order space with the POD basis representing electron wave functions (WFs) guided by the first principles in simulations. To minimize training effort and enhance robustness of POD-Galerkin in larger structures, the quantum element method (QEM) was proposed previously, which partitions nanostructures into generic quantum elements. Larger nanostructures can then be constructed by the trained generic quantum elements, each of which is represented by its POD-Galerkin model. This work investigates QEM-Galerkin thoroughly in multi-element quantum-dot (QD) structures on approaches to further improve training effectiveness and simulation accuracy and efficiency for QEM-Galerkin. To further improve computing speed, POD and Fourier bases for periodic potentials are also examined in QEM-Galerkin simulations. Results indicate that, considering efficiency and accuracy, the POD potential basis is superior to the Fourier potential basis even for periodic potentials. Overall, QEM-Galerkin offers more than a 2-order speedup in computation over direct numerical simulation for multi-element QD structures, and more improvement is observed in a structure comprising more elements.

## 1. Introduction

Nanostructures and nanomaterials have played crucial roles in emerging technologies, including photonics [1-3], electronic devices [4-6], biology and medicine [7,8]. Nanotechnology thus has profound impacts on society today, including computing, communications, sensors, robotics, drugs, medicine, transportations, entertainments, etc. Nanotechnology involves nanoscale or sub-nanoscale materials and devices, including molecular/atomic scales, in which classical approaches are no longer viable. In such a small scale, quantum mechanics are needed to account for lattice formation, electronic band structures, electron transport, electron-photon and electron-phonon interactions, etc. Research regarding nano or molecular scales usually requires solution of the Schrödinger or Schrödinger-like equations subjected to electric field/potentials, defects, photons, phonons, etc. Upon solving the Schrödinger or Schrödinger-like equation, one can determine material properties and device characteristics, including electronic, optical, magnetic and mechanical properties needed for engineering/scientific design.

Currently, one of the most popular tools for *ab initio* simulations of nanoscale or sub-nanoscale structures is Density Functional Theory (DFT). Instead of solving the many-body Schrödinger equations for an *N* electron system, DFT involves solution of *N* single-electron Schrödinger-like equations, also known as the Khon-Sham equations, where the electron-electron interaction is described by an effective potential which is a functional of the electron density. With DFT, one can simulate reaction mechanisms for drugs, charge trapping characteristics based on defects in devices, optical and electrical properties of conducting organic polymers, the molecular dynamics of methanol fuel cells, etc. [9-15]. While this *ab-into* method greatly reduces the computational complexity of many-body Schrödinger equations, it is still computationally intensive, especially periodic nanostructures with defects [10,14,15].

Solution of the partial differential equation (PDE) often involves direct numerical simulation (DNS) or projection of the system onto an *assumed* basis set, such as Fourier basis, wavelets, Legendre polynomials, Bessel functions, Airy functions, Hermite polynomials, etc. DNS approaches are in general computationally intensive for multi-dimensional problems, especially when high resolution is needed in a large domain, due to the required large degree of freedom (DoF). Although the conventional projection-based approaches are able to reduce the DoF, they only work well when the solution is close to the forms of the selected/assumed basis set. The commonly seen examples include Legendre polynomials for solutions near spherical symmetry, Fourier plane waves for periodic structures and Bessel functions for problems having cylindrical symmetry.

Instead of assuming the basis, data-driven reduced order methods, such as Proper Orthogonal Decomposition (POD), Singular Value Decomposition (SVD) and Principal Component Analysis (PCA), have increased in popularity due to their ability to extract an optimal set of basis functions (or modes) from training data to reduce the DoF for the problem of interest [16-20]. In particular, the classical learning algorithms based on the POD basis, that offers the best least squares (LS) fit to the training data, has been implemented in Galerkin projection of the governing PDE in fluids [21-23] and heat transfer [24-26]. Recently, approaches based on POD-Galerkin guided by first principles have been applied to solve electron wave functions (WFs) and their eigenenergies in single-domain nanostructures [27,28]. These studies have shown that with only a few DoF one can predict accurate WFs and eigenenergies of quantum-well (QW) and quantum-dot (QD) structures with learning beyond the training bounds. This is due to the *ab initio* Galerkin projection of the Schrödinger equation, which allows every single calculation in simulation to be guided by first principles to offer the extrapolation ability and improve the accuracy. This is different from physics-informed neural networks (PINNs) for physics simulations [29-31] where physical principles are indeed brought into the training, but no physics guidance is provided during the simulation/prediction.

While the classical physics-informed POD-Galerkin methods discussed above can greatly reduce the DoFs, this approach is not without its drawbacks. Accuracy of data-driven methods for physics simulations is strongly dependent on training thoroughness and accuracy of numerical training data [24, 25, 28]. For large-scale complex nanostructures, this is problematic as computational resources needed in training increase exponentially with the size of the simulation domain and with the DoF of varying parameters. Therefore, DNSs needed to generate high-quality data to include a combination of possible parametric variations may become too extensive in a complex large-domain problem. A solution to ease the problem of the massive data collection and intensive pre-calculations of POD-Galerkin model parameters via the training data is to use the Quantum Element method (QEM) recently proposed in [32, 33]. This method implements domain decomposition to partition large-scale nanostructures into subdomains, each of which is termed a quantum element. POD-Galerkin is thus applied to each of the selected quantum elements. The QEM-POD-Galerkin (or briefly QEM-Galerkin) method not only leads to substantial data reduction in data collection but also minimizes the complexity of parametric variations in training smaller elements. One can then afford to implement a finer mesh in training to improve training data accuracy/quality. These improve training efficiency and resilience of trained POD-Galerkin models to further enhance the prediction accuracy and efficiency.

To make the approach more versatile, the majority of the selected elements should be generic building blocks; namely, once the generic element models are trained and developed for a specific technology or category, these trained element models can be saved in a technology library. These trained elements can then be *stitched* together to form larger and more complex nanostructures in the same technology, using the interior penalty discontinuous Galerkin (IPDG) method [34,35] to enforce the interface continuity. Use of building blocks has been a general practice in modern technologies to improve the cost-effective design/analysis. For example, building blocks like standard cells and functional units have been utilized to facilitate more effective design for VLSI circuits and microprocessors, where complexities of the structures and models are represented by smaller and less-complicated multiple building blocks. This also applies to the concept of materials genome for materials synthesis and design.

The QEM was proposed and validated previously in 1D muti-element QW structures subjected to variation of external electric field [32]. A brief conference study in [33] demonstrated the concept of QEM-Galerkin in small 2D QD structures which focuses on training robust POD modes of elements that are able to predict WFs subjected to electric fields in arbitrary directions and interface variations induced by neighboring elements. The current work investigates ideas to overcome some drawbacks in the previous QEM-Galerkin work and to further enhance its training effectiveness, simulation efficiency and accuracy and extrapolation capability. Only one set of POD modes for each generic element were involved in previous work. In this work, each generic element is projected onto functional spaces constituted by two different basis sets for two distinct physical quantities in *ab initio* QEM-Galerkin simulations (i.e., simulations of multi-quantum-element nanostructures in POD space guided by *ab initio* Galerkin projection). One of the 2 sets is the POD modes for representing the WFs (hereafter named POD WF modes) similar to other quantum POD-related work [27, 28, 32, 33], and the other is introduced to represent the nonuniform potential profile, $U(\vec{r})$, (hereafter named POD $U$ modes) and to speed up computation of the $U(\vec{r})$ projection onto the WF POD space to facilitate faster QEM-Galerkin simulations. Two $U$ basis sets, including POD and Fourier bases, with different approaches are first examined, and the POD $U$ basis with least squares (LS) fitting appears to be an optimal approach according to both efficiency and accuracy. A significant reduction in computing time is demonstrated using the optimal $U$ modes to evaluate the accurate potential projection onto WF POD space in QEM-Galerkin simulation.

The QEM-Galerkin simulation methodology can be applied to effective simulations of periodic nanostructures with or without defects, DFT simulations of supercells, efficient tolerance tests for design of nanostructures influenced by inevitable manufacturing fluctuations, etc. This will become clearer with the demonstrations presented in the following sections. Moreover, due to a clear resemblance between the Helmholtz and Schrödinger equations, QEM-Galerkin can be revised without much difficulty for simulations of photonic crystals, superlattices and metamaterials. The following presentation is organized below. Sec. 2 briefly overviews the single-element and QEM Galerkin methodologies, followed by formulation of a potential basis expansion in QEM-Galerkin. Several approaches of possible basis sets for the spatial potential are discussed. In Sec. 3, selection of an optimal potential basis is presented. Training and demonstrations for multi-element QD structures are performed in terms of computational efficiency and accuracy and the extrapolation capability. In Sec. 4, findings and concepts resulting from this work are discussed, and conclusions are summarized in Sec. 5.

## 2. Quantum-Element Method

### 2.1 Classical POD-Galerkin method for quantum nanostructures

The electron WF $\psi(\vec{r})$ in nanostructures or sub-nanostructures is described by the Schrödinger equation,

$$-\frac{\hbar^2}{2m}\nabla^2\psi(\vec{r}) + U(\vec{r})\psi = E\psi(\vec{r}), \tag{1}$$

where $U$ is the potential, $E$ is the total energy and $m$ is the free electron mass or effective mass for materials or nanostructures in the atomic or nano scale. Instead of solving this equation directly, the physics-informed POD-Galerkin method is applied to solve the problem in a lower-DoF space described by an optimal set of POD WF modes. To generate these modes, one needs to collect training data of WFs from DNSs accounting for parametric variations in the system including, for example, potential and BCs. A set of modes is created to capture the maximum $LS$ information in the collected WF data, using the Fredholm equation of the second kind [17,18],

$$\int_{\Omega'} \boldsymbol{R}(\vec{r},\vec{r}')\eta(\vec{r}')d\vec{r}' = \lambda\eta(\vec{r}), \tag{2}$$

where $\boldsymbol{R}(\vec{r},\vec{r}')$ is the spatial two-point correlation tensor of the collected data given by

$$\boldsymbol{R}(\vec{r},\vec{r}') = \langle \psi(\vec{r}) \otimes \psi(\vec{r}') \rangle, \tag{3}$$

and $\eta$ is a POD WF mode with eigenvalue $\lambda$. The $i$th eigenvalue, $\lambda_i$, represent the mean squared information captured by $\eta_i$. After creating these POD WF modes, one can express the WF as a linear combination of these modes,

$$\psi(\vec{r}) = \sum_{j=1}^{M} a_j \eta_j(\vec{r}), \tag{4}$$

where the weights $a_j$ are the WFs in POD space and $M$ is the number of modes or the DoF used to represent the WFs.

It should be noted that solving (2) directly from discrete data is impractical, especially for multi-dimensional WF data with high resolution. In practice, the method of snapshots [25,36] is used to transform the eigenvalue problem from a discrete spatial domain with dimensions of $N_r \times N_r$ to a sampling domain of size $N_s \times N_s$, where $N_r$ is the number of spatial grid points and $N_s$ is the number of samples. Generally,, $N_s \ll N_r$, and the method of snapshots only generates the first $N_s$ modes. Therefore, it is important ensure that $N_s > M$ and $\lambda_{N_s}$ is many orders smaller than $\lambda_1$.

To close the system, the weights $a_j$ in (4) can be found via the *ab initio* Galerkin projection of the Schrödinger equation onto the POD WF modes. The projection onto the $i$th POD WF mode, $\eta_i$, is given as

$$\int_{\Omega} \nabla\eta_i(\vec{r}) \cdot \frac{\hbar^2}{2m}\nabla\psi(\vec{r})d\Omega + \int_{\Omega} \eta_i(\vec{r})U(\vec{r})\psi(\vec{r})d\Omega - \int_{S} \eta_i(\vec{r})\frac{\hbar^2}{2m}\nabla\psi(\vec{r}) \cdot d\vec{S} = E\int_{\Omega} \eta_i(\vec{r})\psi(\vec{r})d\Omega. \tag{5}$$

Substituting (4) into (5), one obtains an $M \times M$ eigenvalue problem in POD space [27],

$$\boldsymbol{H}_\eta \vec{a} = E\vec{a}, \qquad (6)$$

where $\vec{a} = [a_1, a_2, \ldots, a_M]^T$ and $H_\eta$ is the Hamiltonian in POD space expressed as

$$\boldsymbol{H}_\eta = \boldsymbol{T}_\eta + \boldsymbol{U}_\eta + \boldsymbol{B}_\eta. \qquad (7)$$

The projected Hamiltonian onto POD space includes the projected kinetic energy matrix,

$$T_{\eta\, i,j} = \int_\Omega \nabla \eta_i(\vec{r}) \cdot \frac{\hbar^2}{2m} \nabla \eta_j(\vec{r}) d\Omega, \qquad (8)$$

the projected potential energy matrix,

$$U_{\eta,ij} = \int_\Omega \eta_i(\vec{r}) U(\vec{r}) \psi(\vec{r})\, d\Omega \qquad (9)$$

and the projected boundary energy matrix,

$$B_{\eta,ij} = -\int_s \eta_i(\vec{r}) \frac{\hbar^2}{2m} \nabla \eta_j(\vec{r}) \cdot d\vec{S}. \qquad (10)$$

## 2.2 *Ab initio* multi-element POD-Galerkin (QEM-Galerkin) method

POD-Galerkin described above projects a nanostructure onto one set of POD WF modes that are trained over the entire domain. For large-scale nanostructures with high resolution, the need for a massive amount of training data and the complexity of parametric variations in training may become prohibitive. As discussed in Sec. 1, the multi-element POD-Galerkin method (or QEM-Galerkin) can be applied, and training of each generic element to generate its POD WF modes is performed individually. This offers advantages of more efficient/effective training and more robust modes to achieve better accuracy and efficiency in the prediction. Similar to the single-element approach, with generated POD WF modes for each generic element, *ab initio* Galerkin projection of the *p*th element onto the Schrödinger equation is performed to derive a set of equations for $\vec{a}_p$. (5) is then revised for the *p*th element, coupled with the (*q*th) neighboring elements,

$$\int_{\Omega_p} \nabla \eta_{p,i} \cdot \frac{\hbar^2}{2m} \nabla \psi_p d\Omega - \sum_{q=1,q\neq p}^{N_{el}} \oint_{S_{pq}} \left[ \left[\!\left[\frac{\hbar^2}{2m}\psi\right]\!\right]_{pq} \langle \nabla \eta_i \rangle_{pq} + \langle \frac{\hbar^2}{2m} \nabla \psi \rangle_{pq} [\![\eta_i]\!]_{pq} \right] \cdot d\vec{S} + \int_{\Omega_p} \eta_{p,i} U \psi_p d\Omega$$

$$-\mu \sum_{q=1,q\neq p}^{N_{el}} \oint_{S_{pq}} \left[\!\left[\frac{\hbar^2}{2m}\psi\right]\!\right]_{pq} [\![\eta_i]\!]_{pq} dS = E \int_{\Omega_p} \eta_{p,i}\, \psi_p d\Omega, \qquad (11)$$

where $N_{el}$ is the number of elements in the entire nanostructure, the IPDG method is applied to the surface integrals to enforce the continuity at element interfaces, and $\mu$ is the penalty parameter defined as $\frac{N_\mu}{dr}$ with $dr$ being the local mesh size at the interface and $N_\mu$ as the non-unit penalty number. Additionally, $[\![*]\!]_{pq}$ and $\langle * \rangle_{pq}$ are the difference and average operators, respectively, across the interface of the *p*th and *q*th elements. Accuracy of the predicted solution may be influenced by the selected value of $N_\mu$.

Using (4) in (11) for each element, a Hamiltonian equation for the *p*th element in POD space can be expressed as,

$$\sum_{j=1}^{M_p} \left(T_{\eta_p,ij} + U_{\eta_p,ij}\right) a_{p,j} + \sum_{q=1,q\neq p}^{N_{el}} \sum_{j=1}^{M_p} B_{p,pq,ij} a_{p,j} + \sum_{q=1,q\neq p}^{N_{el}} \sum_{j=1}^{M_q} B_{pq,,ij} a_{p,j} = E a_{p,i}, \tag{12}$$

where $M_p$ and $M_q$ are the numbers of modes for the $p$th and $q$th elements, respectively. Additionally, $T_{\eta_p,ij}$ is the projected interior kinetic energy matrix for the $p$th element,

$$T_{\eta_p,ij} = \int_\Omega \nabla \eta_{p,i} \frac{\hbar^2}{2m_p} \nabla \eta_{p,j} d\Omega, \tag{13}$$

$U_{\eta_p,ij}$ is the projected potential energy matrix,

$$U_{\eta_p,ij} = \int_{\Omega_p} \eta_{p,i} U(\vec{r}) \eta_{p,j} d\Omega, \tag{14}$$

$B_{p,pq,ij}$ is the projected off-diagonal boundary kinetic energy matrix,

$$B_{pq} = -\frac{1}{2} \int_{S_{pq}} \left(\nabla \eta_{p,i}\right) \eta_{p,j} + \eta_{p,i} \left(\nabla \eta_{p,j}\right) \cdot d\vec{S} + \mu \int_{S_{pq}} \frac{\hbar^2}{2m} \eta_{p,i} \eta_{p,j} dS, \tag{15}$$

and the projected off-diagonal boundary kinetic energy matrix, $B_{pq,,ij}$, is defined as

$$B_{pq,ij} = \frac{1}{2} \int_{S_{pq}} \left(\nabla \eta_{p,i}\right) \eta_{q,j} - \eta_{p,i} \left(\nabla \eta_{q,j}\right) \cdot d\vec{S} - \mu \int_{S_{pq}} \frac{\hbar^2}{2m} \eta_{p,i} \eta_{q,j} dS. \tag{16}$$

The QEM-Galerkin Hamiltonian equation thus from (12) becomes

$$\begin{bmatrix} \mathbf{H_1} & \mathbf{H_{1,2}} & \cdots & \mathbf{H_{1,q}} & \cdots & \mathbf{H_{1,N_{el}-1}} & \mathbf{H_{1,N_{el}}} \\ \mathbf{H_{2,1}} & \mathbf{H_2} & \cdots & \mathbf{H_{2,q}} & \cdots & \mathbf{H_{2,N_{el}-1}} & \mathbf{H_{2,N_{el}}} \\ \vdots & \vdots & \ddots & \vdots & \reflectbox{$\ddots$} & \vdots & \vdots \\ \vdots & \vdots & \cdots & \mathbf{H_p} & \cdots & \vdots & \vdots \\ \vdots & \vdots & \reflectbox{$\ddots$} & \vdots & \ddots & \vdots & \vdots \\ \mathbf{H_{N_{el}-1,1}} & \mathbf{H_{N_{el}-1,2}} & \cdots & \mathbf{H_{N_{el}-1,q}} & \cdots & \mathbf{H_{N_{el}-1}} & \mathbf{H_{N_{el}-1,N_{el}}} \\ \mathbf{H_{N_{el},1}} & \mathbf{H_{N_{el},2}} & \cdots & \mathbf{H_{N_{el},q}} & \cdots & \mathbf{H_{N_{el},N_{el}-1}} & \mathbf{H_{N_{el}}} \end{bmatrix} \begin{bmatrix} \vec{a}_1 \\ \vec{a}_2 \\ \vdots \\ \vec{a}_q \\ \vdots \\ \vec{a}_{N_{el}-1} \\ \vec{a}_{N_{el}} \end{bmatrix} = E \begin{bmatrix} \vec{a}_1 \\ \vec{a}_2 \\ \vdots \\ \vec{a}_q \\ \vdots \\ \vec{a}_{N_{el}-1} \\ \vec{a}_{N_{el}} \end{bmatrix}, \tag{17}$$

where the diagonal block entries $H_p$ are given as

$$H_{p,ij} = \sum_{j=1}^{M_p} \left(T_{\eta_p,ij} + U_{\eta_p,ij}\right) + \sum_{q=1,q\neq p}^{N_{el}} \sum_{j=1}^{M_p} B_{p,pq,ij}, \tag{18}$$

and the off-diagonal entries are expressed as

$$H_{pq,ij} = \sum_{j=1}^{M_q} B_{pq,,ij} \text{ when } p \neq q. \tag{19}$$

Note that $M_p$ is smaller than the total number of snapshots ($N_{s,p}$) of the training data collected from the $p$th element. After solving (17), the WF in the $p$th element can expressed as a linear combination of the modes with weights $\vec{a}_p$ as,

$$\psi_p(\vec{r}) = \sum_{j=1}^{M_{p'}} a_{p,j}\eta_{p,j}(\vec{r}) = \vec{a}_p^T \cdot \vec{\eta}_p, \qquad \text{for } p = 1 \text{ to } N_{el} \tag{20}$$

with $M_{p'} \leq M_p$, $M_{p'}$ is the selected number of modes for the $p$th element in post processing.

There are four major steps in this approach from data collection to the WF prediction. The first step is the training, involving data collection and mode generation. The second step is to evaluate all QEM-Galerkin model parameters to construct the QEM-Galerkin Hamiltonian matrix given in (17). These modes and model parameters are precalculated and saved in the technology library except for the potential entries $U_{\eta_p,ij}$ of (18) expressed in (14) due to the nonuniform potential $U(\vec{r})$ that is unknown during the training. $U_{\eta_p,ij}$ thus needs to be computed during simulation if $U(\vec{r})$ is nonuniform. A more detailed discussion is given in Sec. 2.3. The third is performing QEM-Galerkin simulation in POD space to solve $\vec{a}_p$ from (17). The last step is to calculate $\psi_p(\vec{r})$ in (20) after solving $\vec{a}_p$, which is referred to as post-processing. To avoid direct integral of (14) involving the nonuniform $U(\vec{r})$ during QEM-Galerkin simulation, projection-based approaches for $U(\vec{r})$ are presented below to minimize this computational burden.

## 2.3 Effective evaluation of potential during QEM-Galerkin simulation

When performing the quantum POD-Galerkin simulation, it is necessary to project the spatial potential $U(\vec{r})$ onto the POD space. This potential calculation in POD space involves a time-consuming integral of $U(\vec{r})$ with POD WF modes in (14) real space. In previous investigations of single- or multi-element quantum POD-Galerkin methods [27, 28, 32, 33], spatial potential was uniform, linear, or piece-wise linear, where $U_{\eta_p,ij}$ in (14) could be computed efficiently in simulation due to linearity with pre-calculated data. In realistic cases, the potential variation may be nonuniform, prohibiting the pre-calculations. To improve the computationally intensive potential calculation in POD space during the QEM-Galerkin simulation, the potential $U(\vec{r})$ is projected onto a set of potential basis functions $\tilde{\eta}_U$ (or $U$ modes),

$$U(\vec{r}) = \sum_{n=1}^{M_U} c_n \tilde{\eta}_{U,n}, \tag{21}$$

where the coefficients $c_n$ are the $U$ projections onto the $U$ modes $\tilde{\eta}_{U,n}$, and $M_U$ is the total number of $U$ modes. Using (21), potential energy in WF POD space for the $p$th element in (14) becomes

$$U_{\eta_p,ij} = \sum_{n=1}^{M_U} c_n \int_{\Omega_p} \eta_{p,i}\, \tilde{\eta}_{U,n}\, \eta_{p,j}\, d\Omega, \tag{22}$$

where the integrals can be pre-calculated to speed up QEM-Galerkin simulation once $\eta(\vec{r})$ and $\tilde{\eta}_U(\vec{r})$ are known. An optimal set of $\tilde{\eta}_U(\vec{r})$ that offers fast convergence in (21) is desired to minimize the computing time during simulation. In this work, both POD and Fourier $U$ modes are applied to examine their computational efficiency and accuracy. For the Fourier basis, which seems to be a natural choice to represent periodic potentials, efficient Fast Fourier Transform (FFT) is applied. For the POD $U$ modes, 2 different approaches are presented below.

When using POD $U$ modes, the modes need to be computed, similarly to how the POD WF modes are generated, based on (2) using the method of snapshots. The WF data in the 2-point correlation tensor $R(\vec{r},\vec{r}')$ is thus replaced by the potential data. Data collection of $U(\vec{r})$ to generate POD $U$ modes is presented below in Sec. 3.2. When using POD $U$ modes, $[c_1\ c_2\ ...\ c_n\ ...\ c_{M_U}]$ in (21) can be evaluated either from direct projections of $U(\vec{r})$ onto each of the POD $U$ modes for $n = 1$ to $M_U$,

$$c_n = \int_\Omega U(\vec{r})\,\tilde{\eta}_{Un}(\vec{r})\,d\Omega, \tag{23}$$

or via *LS* fitting

$$[\tilde{\eta}_{U1}\ \tilde{\eta}_{U2}\ ...\ \tilde{\eta}_{Un}\ ...\ \tilde{\eta}_{UM_U}][c_1\ c_2\ ...\ c_n\ ...\ c_{M_U}]^T = U(\vec{r}). \tag{24}$$

In (24), $c_n$ are found by the Moore–Penrose inverse (i.e, pseudoinverse) that leads to the best *LS* fit to data of the spatial potential variation. The number of selected spatial grid points $N_{Ur}$ for $\tilde{\eta}_{Un}(\vec{r})$ in (24) should be considerably greater than $M_U$ in (21) to ensure an accurate *LS* fitting, i.e., $N_{Ur} \gg M_U$. In this study, we choose $N_{Ur} \approx 6M_U$ with randomly selected grid points. Use of a larger $N_{Ur}$ will improve the accuracy but require more computational time during QEM-Galerkin simulation. Approaches using POD and Fourier $U$ modes are examined first in Sec. 3.2. Based on consideration of efficiency and accuracy, the optimal approach is applied in demonstrations of QEM-Galerkin in Sec. 3.3.

## 3. Training and Demonstrations

### 3.1 Quantum dot potential in test structures

Demonstrations of QEM-Galerkin are carried out in two multi-element QD test structures, Structure A with 4 × 4 elements and Structure B with 20 × 20 elements, as shown in Table 1. For each of these two structures, one generic QD element with a size of 16 nm × 16 nm is trained and used to construct the entire multi-element structure. To generate a set of POD WF modes for each generic element to represent the WFs responding to variations in the QD potential, the generic element is trained in a domain with 2 × 2 elements. The QD potential in each generic element is modeled as an upside-down Gaussian function centered at $(x_o, y_o)$ with an amplitude of $A$ and a standard deviation of $\sigma$,

$$f(x,y) = A\left(1 - e^{-\frac{(x-x_0)^2+(y-y_0)^2}{2\sigma^2}}\right) \tag{25}$$

with $A$ = 0.8 eV. As shown in Table 1, 50 samples of potentials over a training domain of 2 × 2 elements are applied in DNSs. For Structure A, $\sigma$ varies uniformly between 5nm and 6.5 nm with $(x_o, y_o)$ fixed at (0,0) in DNSs of the Schrodinger equation to collect WF training data accounting for the potential variation. For Structure B, 3 parameters randomly change within their bounds independently, as specified in Table 1; namely, $\sigma$ varies between 5nm and 6.5 nm, and both $x_o$ and $y_o$ vary between -1 nm to +1 nm. In both training and demonstration, a periodic potential for the entire simulation domain (instead of each element) is enforced, and the potential given in (25) is extended across each element interface (including the periodic boundary of the entire domain) to the neighboring elements. The total potential over the entire domain is thus constructed by the superposition of all element potentials.

Table 1 Variation of training potentials for generation of POD WF modes

| Test Structure | No. of elements in training | Variations of 50 training potential for WF data collection | | |
|---|---|---|---|---|
| | | $\sigma$ (nm) | $x_o$ (nm) | $y_o$ (nm) |
| A (4 × 4 elements) | 2 × 2 | [5, 6.5] | 0 | 0 |
| B (20 × 20 elements) | 2 × 2 | [5, 6.5] | [-1, +1] | [-1, +1] |

Accuracy of QEM-Galerkin models strongly depends on the training mesh resolution in DNSs. A fine mesh of $200 \times 200$ is thus implemented in each generic element in all cases in training to ensure good accuracy of the numerical training data. The resolution of WFs in post processing via (20) is identical to the mesh resolution of the training data. If fine resolution is not needed, coarser resolution can be applied in (20) to speed up the post processing.

**3.2 Selection of potential basis functions ($U$ modes)**

Before demonstrating QEM-Galerkin in multi-element structures, approaches using POD and Fourier $U$ modes for evaluating the $U$ projections $c_n$ in (21) are examined first to select an optimal approach. When using Fourier $U$ modes, the FFT is applied to evaluate $c_n$ with pre-evaluated integrals in (22). When applying POD $U$ modes, training potential data is collected from a training domain of $5 \times 5$. Note that selection of the number of elements in training does not affect accuracy and resilience of POD $U$ modes as long as a large number of data samples are collected to account for enough interior and boundary potential variations. One set of POD $U$ modes is generated for each generic element, including Element A for Structure A or Element B for Structure B. The former accounts for random variation in $\sigma$ and the latter for random variations in both $\sigma$ and $(x_o, y_o)$ within the same training bounds for generating POD WF modes given in Table 1. To generate POD $U$ modes for each generic element, 60 random samples of potential data are collected from the $5 \times 5$ training domain in addition to the fifty $2 \times 2$ potential samples used in DNSs for collecting WF data (see Table 1). Therefore, the total number of potential snapshots for each generic element equals $60 \times 5 \times 5 + 50 \times 2 \times 2 = 1700$.

One of potential profiles in the training domain of $5 \times 5$ elements is shown in Fig. 1(a), and the eigenvalue spectrums for both generic elements are illustrated in Fig. 1(b). Because variation embedded in potential data for Element A is less complicated than that in Element B, Element A's eigenvalue declines more rapidly, which should result in a faster convergence of (21) for $U(\vec{r})$. For the Moore-Penrose inverse (LS fitting) in (24), the number of spatial grid points $N_{Ur} = 132$ for $\tilde{\eta}_{Un}(\vec{r})$ is selected for both generic elements even though Element B is subjected to more complicated potential variation.

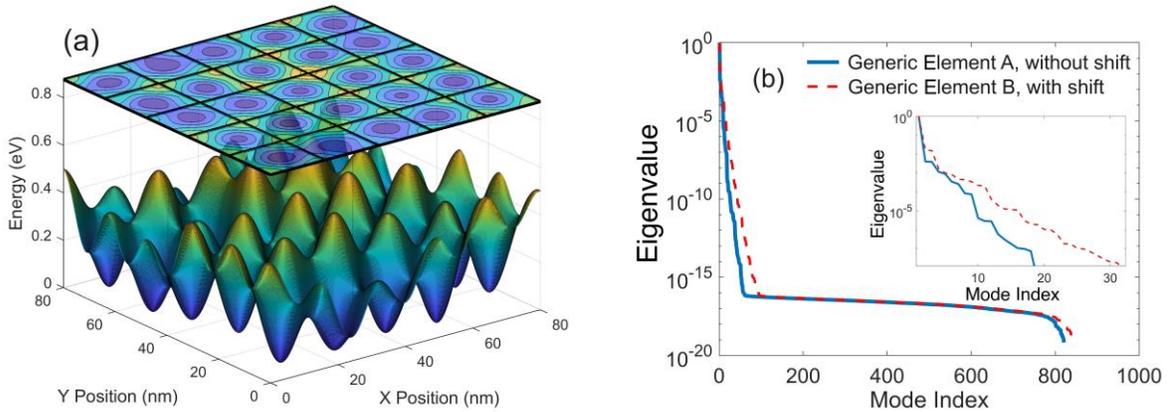

Fig. 1. (a) A potential profile in the domain with $5 \times 5$ elements for training of POD $U$ modes. (b) Eigenvalue spectrum of potential data from each generic element with a zoom-in inset.

While training of POD $U$ modes for each generic element is carried out in a periodic domain with $5 \times 5$ elements, validation of $U$-mode approaches for each element is performed in its test structure. Accuracy of WFs predicted by QEM-Galerkin with respect to DNS's results is strongly influenced by the consistency of potential profiles between both QEM-Galerkin and DNS approaches. Highly accurate $U(\vec{r})$ represented by the $U$ projections $c_n$ in (21) is thus desired in QEM-Galerkin to minimize the WF error predicted by QEM-Galerkin. The number of POD $U$ modes for either the *LS* fitting in (24) or POD projection in (23) (see the discussion in Sec. 2.3) is thus selected to reach an *LSE* of $U(\vec{r})$ near or below 0.1%, as shown in Tables 2 and 3 for Structure A and B, respectively. An *LSE* near 0.55% - 0.6% is achieved by 15 POD $U$ modes for Element A in Structure A using either (23) or (24) to evaluate $c_n$ in (21). Due to more complicated potential variation in Structure B, an *LSE* near 0.108% - 0.129% is reached with 21 $U$ modes. For each test structure between the 2 approaches of (23) and (24) whose *LS* errors are close, the *LS* fitting approach in (24) is 13 (3.51/0.273) and 17 (5.31/0.319) times faster for Structure A and Structure B, respectively, than the direct POD projection in (23).

When using Fourier $U$ modes for the entire domain, with $c_n$ in (21) evaluated using FFT, Table 2 shows that it requires 225 (or 15 × 15) or 256 (16 × 16) modes in Structure A to reach a similar *LSE* to the POD $U$ mode approaches. Constructing (22) using Fourier $U$ modes is around 20 to 22 (5.52/0.273 to 6.27/0.273) times slower than using 15 POD $U$ modes with *LS* fitting. For Structure B, 196 Fourier $U$ modes for the entire domain are needed to reach an *LSE* similar to the POD $U$ mode approaches and this Fourier approach is around 15 (4.68/0.319) times slower than the *LS* fitting approach that uses only 21 POD $U$ modes.

Table 2 Performance of $U$ projections in Structure A

| Method using $U$ modes | | *LSE* of $U(\vec{r})$ (%) | No. of $U$ modes | Time per element (ms) | Improved efficiency |
|---|---|---|---|---|---|
| POD | *LS* fitting (Element A) | 0.060 | 15 | 0.273 | 386 |
| | Projection (Element A) | 0.0552 | 15 | 3.51 | 30 |
| Fourier basis | Entire domain | 0.0728 | 15×15 | 5.52 | 19.1 |
| | | 0.0487 | 16×16 | 6.27 | 16.8 |
| | Element A | 3.39 | 15×15 | 4.56 | 23.1 |
| Direct integral of (9) | | - | - | 105.3 | 1 |

Table 3 Performance of $U$ projections in Structure B

| Method using $U$ modes | | *LSE* of $U(\vec{r})$ (%) | No. of $U$ modes | Time per element (ms) | Improved efficiency |
|---|---|---|---|---|---|
| POD | *LS* fitting (Element B) | 0.129 | 21 | 0.319 | 362 |
| | Projection (Element B) | 0.108 | 21 | 5.31 | 21.8 |
| Fourier basis | Entire domain | 0.151 | 13×13 | 4.36 | 26.5 |
| | | 0.101 | 14×14 | 4.68 | 24.7 |
| | | 0.0771 | 15×15 | 5.56 | 20.8 |
| | Element B | 2.58 | 25×25 | 11.2 | 10.3 |
| Direct integral of (9) | | - | - | 115.6 | 1 |

Note that mode generation/training is not needed for Fourier $U$ modes, and $c_n$ in (21) are evaluated once for the entire test structure using FFT. By comparison, POD $U$ modes are trained and generated for each generic element; hence, the $c_n$ are computed distinctly for each element. The element approach using POD $U$ modes is more flexible and the pre-calculated data can be applied to any periodic or nonperiodic structure within or near the training bounds. Fourier $U$ modes can also be applied to generate a generic element which however will be only accurate for periodic elements. When using 225 $U$ modes in Structure A for both Fourier approaches, as displayed in Table 2, the Fourier approach based on the pre-calculated generic element is 21% faster than that for the Fourier expansion in the entire structure; however, the *LSE* of the approach using the generic element is as high as 3.39%. To improve the accuracy, 625 modes are used in the Fourier generic element for Structure B shown in Table 3. Its *LSE* is reduced but still as high as 2.8%, and the computational time becomes double of what the other Fourier approach needs. The higher *LSE* resulting from the Fourier generic element is caused by the non-periodic elements in Structures A and B even though the entire structures are periodic.

When comparing computational times of the approaches using $U$ modes to evaluate $U_{\eta_p,ij}$ in (22) to the direct numerical integral of (14), Tables 2 and 3 show that the *LS* fitting approach offers an improvement of 386 and 362 times for Structures A and B, respectively. Improvements of all other $U$-mode approaches over the direction integral are only 10 to 30 times. Considering efficiency and accuracy among all $U$-mode approaches, the *LS* fitting for the POD $U$ modes is clearly superior to all others. In the following demonstration of QEM-Galerkin, the *LS* fitting for the POD $U$ modes is applied. That is, the expansion of $U(\vec{r})$ in (21) is applied in POD space to evaluate $U_{\eta_p,ij}$ in (14) with the coefficients $c_n$ in (22) determined by LS fitting using (24) during QEM-Galerkin simulations.

### 3.3 Demonstrations of QEM-Galerkin methodology

There are two sets of POD modes for each trained generic element, including POD $U$ modes and POD WF modes. Training of POD $U$ modes has been detailed in Sec. 3.2. For training the POD WF modes, potential variation applied in DNSs to collect WF training data is shown in (25) and Table 1 of Sec. 3.1 for both test structures. Effective training of POD WF modes relies on good quality of collected WF data from DNSs. Our thorough investigation has found that training data quality is strongly influenced by the following factors

(a) Numerical accuracy of WF data. A fine mesh in DNSs is thus needed to ensure highly accurate training data, and a mesh of $200 \times 200$ is taken for each generic element for the 2 test structures.
(b) Training parameters that need to cover as much as possible the variation within the training bounds, especially for cases involving multi-parameters, such as $\sigma$ and $(x_o, y_o)$ in Structure B.
(c) The number of QSs included in the collected WFs. It is found that collecting WFs in more QSs to generate POD WF modes than what is needed in (17) significantly enhances resilience and quality of the generated POD WF modes. Guided by first principles via *ab initio* Galerkin projection, these improved modes lead to better prediction accuracy within and beyond training bounds.

In addition to the training factors that affect the quality and robustness of the trained POD modes, prediction accuracy can also be improved when using (a) more modes ($M_{p'}$) for post processing in (20) for each element and/or (b) a larger dimension $M_p$ for $\mathbf{H}_p$ for each element in (17) during simulation than what is needed in (20). The above factors in training, simulation and post processing are considered and examined to observe the influences of these factors. Briefly, these are summarized as follows.

The demonstrations below for both test structures predict WFs in the first 15 QSs. For Structure A, training data for WFs are collected from the first 15 states. Using a dimension of $20 \times 20$ for the POD Hamiltonian in each element (i.e., $M_p = 20$) in (17) with spatial potential variation in simulation within the training bounds, the first demonstration shows an accurate prediction of all WFs with a small number of POD WF modes. When spatial potential in simulation is beyond the training bounds in Structure A, the quality of the same set of POD WF modes becomes insufficient, which deteriorates the prediction accuracy for many states even using a larger number of POD WF modes. Without retraining the POD WF modes to cover enough potential variation, it is demonstrated that the prediction accuracy in this extrapolation case using the same number of POD WF modes in post processing can be evidently improved by increasing $M_p$ in the QEM-Galerkin simulation to 30. As shown in Table 1, 50 samples of training potential profiles in Elements A and B are used for Structures A and B, respectively. The former involves variation in only one DoF in $\sigma$ while the latter includes variations in 3 DOF in $\sigma$, $x_o$ and $y_0$. The parameter variations embedded in the training data from Element B are not thorough enough compared to Element A. Instead of including more training data with more thorough variations in 3 DoF ($\sigma$, $x_o$ and $y_0$) that would require considerably more training time, based on Item (c), WFs are collected in 30 QSs (more than 15 states needed in the prediction) to improve the data quality and robustness of the trained POD WF modes.

**Structure A: 4 × 4 elements**

The first test structure is constructed from a 4 × 4 grid using the one trained element, Element A, to predict the first 15 QS WFs. To generate POD WF modes for this generic element, 50 samples of WF data for the first 15 states are collected from DNSs of a training domain with 2 × 2 elements subjected to potential variation described in Table 1. WF data collected from the 2 × 2 elements of the training domain are stacked together in the method of snapshots (a total of 15 × 50 × 2 × 2 snapshots) to generate one generic set of POD modes. One of the training potential profiles in the 2 × 2 training domain is shown in Fig. 2(a), and the eigenvalue of the WF data is given in Fig. 2(b) in descending order. Each eigenvalue represents the information on WF squared $|\psi|^2$ captured by its mode. Because of simple potential variation in training Element A, the eigenvalue decreases rather quickly, as observed in Fig. 2(b), where it drops 3 to 4 orders from the first to the 12th - 15th mode. This indicates that WFs in most states in each element should be well presented by (20) using 12 to 15 modes if the training data quality and accuracy of potential estimated by the POD $U$ modes are sufficient. Also, as discussed above, a higher dimension $M_p$ for $\mathbf{H}_p$ in (17) in QEM-Galerkin simulation can be used to improve the prediction accuracy.

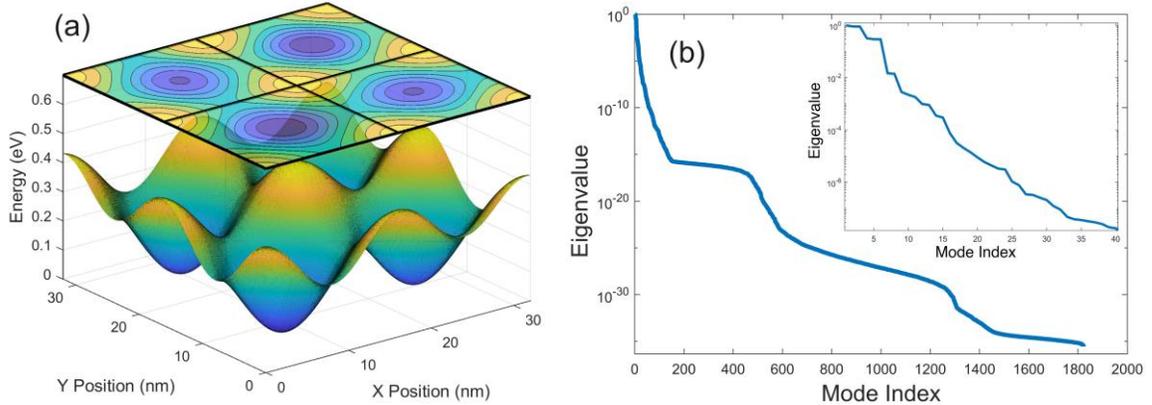

Fig. 2. (a) An example of a potential in the training domain with 2 × 2 elements. (b) POD eigenvalue spectrum derived from 3,000 snapshots of WF data for the generic element.

To verify the QEM-Galerkin methodology for Structure A, a random $\sigma$ value within the training bounds is assigned in each element to determine the potential seen in Fig. 3. WFs in 15 states are solved from the QEM-Galerkin Hamiltonian in (17) for Structure A with a dimension of 20 × 20 ($M_p = 20$) for each element Hamiltonian $\mathbf{H}_p$ in (17). Profiles of $|\psi|^2$ in several QSs predicted by QEM-Galerkin using $N_\mu = 0.01$ are illustrated in Fig. 4, compared against DNS results. Note that the WF of each state in the multi-element structure may be confined to one or a few elements. Some POD WF modes may have nearly no contribution to certain states in some elements, i.e., their coefficients $a_{p,j}$ solved from (17) may be nearly zero and thus ignored in post processing via (20). Hence, the number of modes for each QS indicated in Fig. 4 is the maximum $M_{p'}$ in (20) (hereafter referred to as $M_{el,m}$) over all elements, $p = 1$ to 16, and a different number of modes is used in each element for each QS. As seen in Fig. 4, the WF in QS 1 can be well represented by $M_{el,m} = 5$ while in higher QSs, more modes are needed.

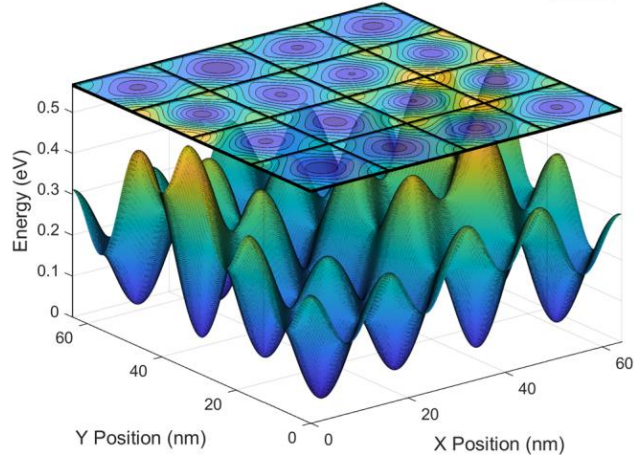

Fig. 3. Potential profile in the test structure with $4 \times 4$ generic elements.

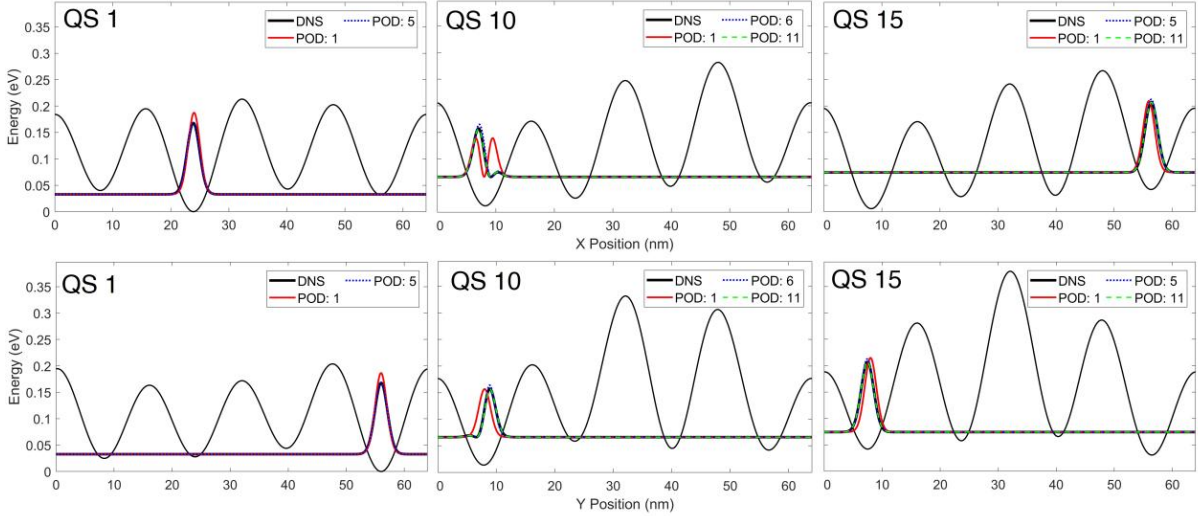

Fig. 4. Profiles of $|\psi|^2$ of QSs 1, 10 and 15, together with the potential profiles, along $x$ and $y$ through the maximum $|\psi|^2$ using $N_\mu = 0.01$. Each legend shows the maximum number of modes (i.e., $M_{el,m}$) in each of the 16 elements in post-processing.

The *LSEs* over all 15 QSs are illustrated in Fig. 5(a) in terms of $M_{el,m}$. The total DoF (i.e., the total number of modes) needed over all 15 states in the entire structure needed in post processing is given in Fig 5(b) as a function of $M_{el,m}$. As shown in Fig. 5(a), to reach an *LSE* below 2% for QSs 1, 2, 6 and 11, it only needs $M_{el,m} = 5$, which leads to the DoF of 75. If an *LSE* below 1% is required for all these states, $M_{el,m} = 6$ with DoF = 90 is needed. For all 15 QSs to reach an *LSE* near or below 2%, $M_{el,m}$ = 10 and DoF = 149. For all 15 QSs to reach an *LSE* near or below 1%, one would involve $M_{el,m}$ = 14 with DoF = 197. Compared to DNS, QEM-Galerkin in this structure offers a reduction in DoF by more than 3 orders of magnitude. Fig. 6 also includes the eigenenergy errors predicted by QEM-Galerkin in the first 15 states, together with the eigenenergies estimated from DNS. The errors in all 15 QSs are very small with a maximum below 0.27% (or 0.105 meV).

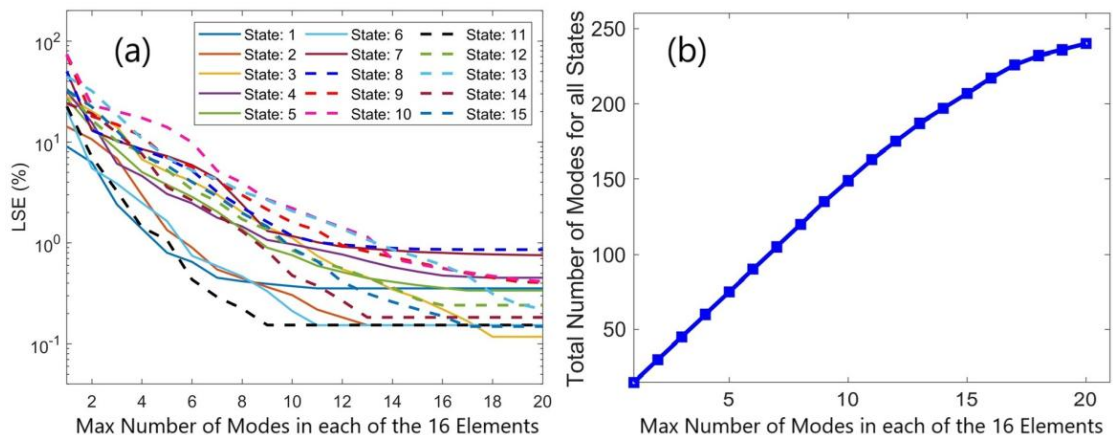

Fig. 5. (a) *LSE* of QSs 1-15 using $N_\mu = 0.01$ as a function of the maximum number of modes ($M_{el,m}$) in each of the 16 elements. (b) Total number of modes (or DoF) used for all 15 states in post processing for the entire test structure.

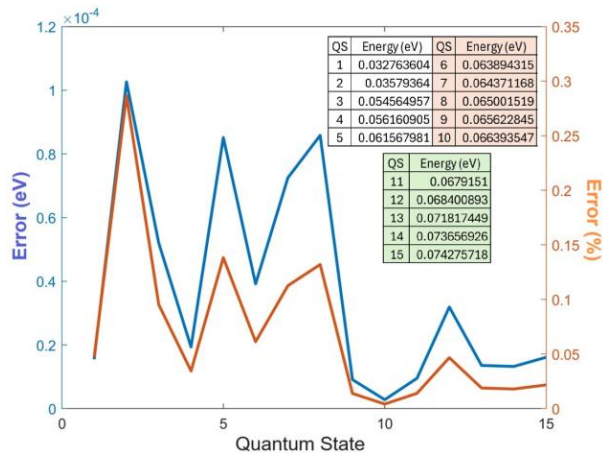

Fig. 6. Error of the eigenenergy for QSs 1-15 in eV and percentage with respect to the minimum potential energy. The inset table shows the QS energies.

To observe the effect of the penalty number $N_\mu$, Figs. 7(a) and 7(b) also illustrate the *LSEs* for QSs 1-15 using higher values of $N_\mu$. For $N_\mu$ varying from 0.01, 1 to 10 shown in Figs. 5(a), 7(a) and 7(b), it is observed that the *LSE* in all states remains nearly unchanged for $M_{el,m}$ < 5. When using $M_{el,m}$ > 5, the *LSE* for the same number of modes increases as $N_\mu$ increases for all QSs but QSs 2 and 11 and 15. In general, when higher accuracy is needed, a smaller value of $N_\mu$ is preferred. However, it is interesting that *LSEs* of WFs in QSs 2, 11 and 15 remain small and nearly unchanged for $N_\mu$ varying from 0.01 to 10 for $M_{el,max} \leq$ 20 in this structure.

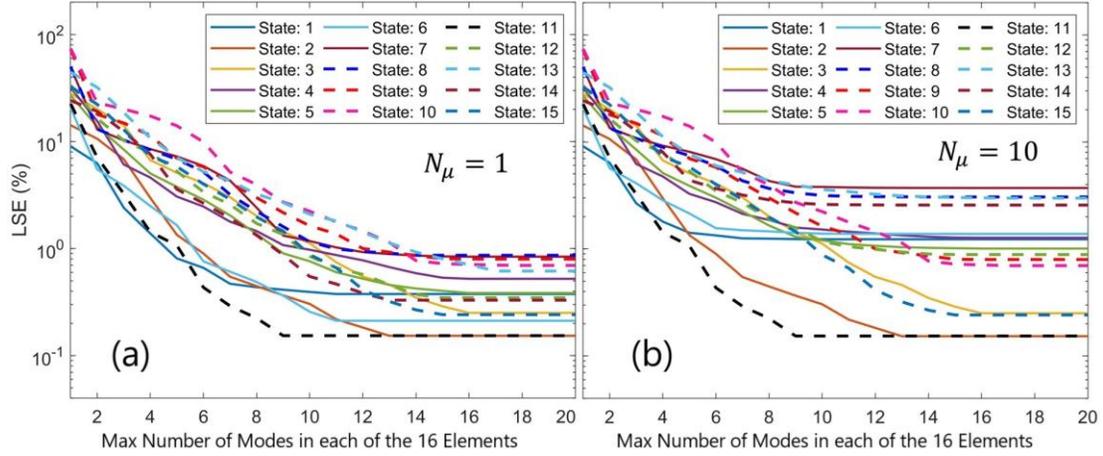

Fig. 7. *LSEs* of QSs 1-15 as functions of the maximum number of modes ($M_{el,m}$) in each of the 16 elements using (a) $N_\mu = 1$ and (b) $N_\mu = 10$.

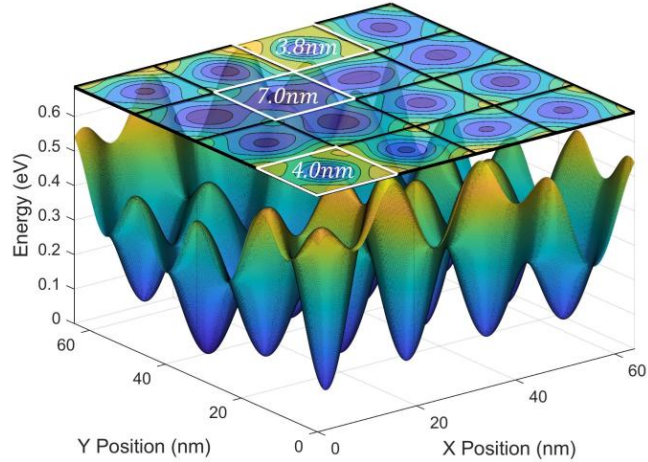

Fig. 8. Potential profile in Structure A for the extrapolation demonstration. As indicated, 3.8 nm, 4nm and 7 nm for $\sigma$ of the potentials in 3 of 16 elements are outside the training range for $\sigma$ (see Table 1). In all other elements, $\sigma$ is assigned randomly within the training bounds.

To examine the learning ability of QEM-Galerkin, the next demonstration is extended beyond the training range for Structure A, as shown in Fig. 8, where potentials in 3 of the elements specified by (25) with their $\sigma$ values outside the training bounds. When using the same set of POD modes in this extrapolation case, the training is not complete (leading to poor training data quality). Thus, the *LSE* is expected to be considerably larger in this case presented in Fig. 9 than in the interpolation case in Figs. 4-6, where the same dimension ($M_p = 20$) in (17) is used in both cases. Fig. 9(a) illustrates the *LSEs* of WFs in QSs 1 - 15, and the total DoF needed in post processing is given in Fig 9(b). The *LSEs* of WFs are relatively large compared to those for the interpolation case in Fig. 5(a). However, good accuracy still can be achieved for most states using a small number of modes. For example, an *LSE* near or below 2% with $M_{el,m} \geq 13$ is observed in Fig. 9(a) (DoF = 184 in Fig. 9(b)) for all states except for QSs 9, 11, 13 and 15, and *LSEs* of WFs in QSs 1, 8, 10 and 14 are below 1% with just 6 or 7 modes. Somehow, *LSEs* in QSs 9, 11, 13 and 15 are considerably worse in this extrapolation case; they are equal to 7.92%, 8.3%, 6.58% and 7.15% when $M_{el,m}$ = 12 and 5.77%, 6.5%, 4.18% and 4.78% when using $M_{el,m}$ = 20, respectively.

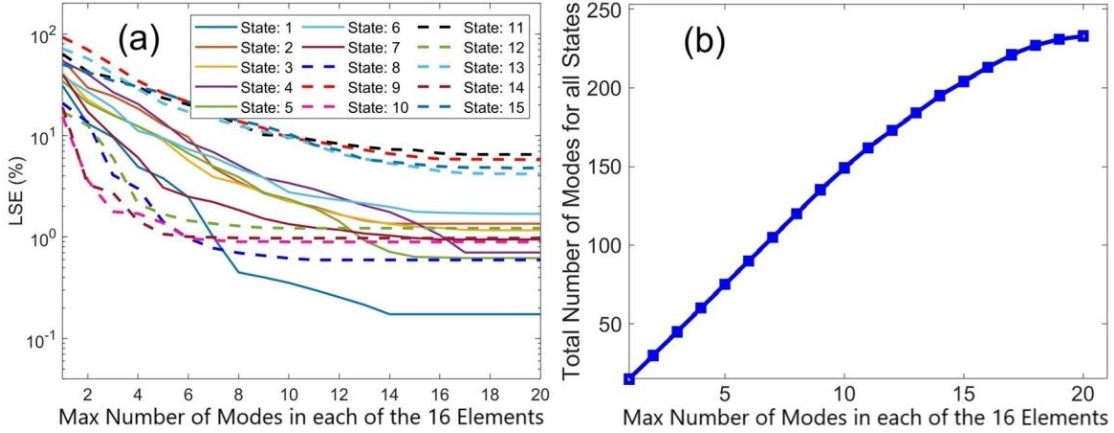

Fig. 9 (a) *LSE* of QSs 1-15 $N_\mu = 0.01$ for the extrapolation case as a function of the maximum number of modes ($M_{el,m}$) in each of the 16 elements. (b) Total number of modes (or DoF) used for all 15 states in post processing. $M_p$ = 20 in (17) for each generic element are applied in QEM-Galerkin simulation.

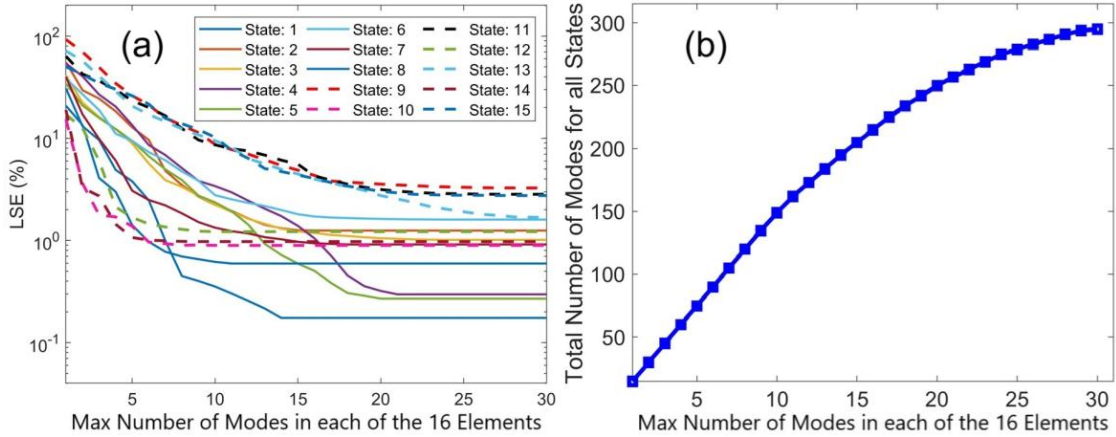

Fig. 10 (a) *LSE* of QSs 1-15 $N_\mu = 0.01$ for the extrapolation case as a function of the maximum number of modes ($M_{el,m}$) in each the 16 elements. (b) Total number of modes (or DoF) used for all 15 states in post processing. $M_p$ = 30 in (17) for each generic element are applied in QEM-Galerkin simulation.

As discussed above, one could increase the dimension $M_p$ in the Hamiltonian $\mathbf{H}_p$ in (17) for each element in simulation to improve the accuracy. When implementing a 30 × 30 Hamiltonian for $\mathbf{H}_p$ (using the same set of POD WF modes trained by WFs in the first 15 states), significant improvement in QEM-Galerkin is observed in Fig. 10(a) using the same number of modes for the case with the 20 × 20 Hamiltonian for $\mathbf{H}_p$. However, a slightly larger DoF shown in Fig. 10(b) is needed in post processing. For example, when using $M_{el,m}$ = 12, the *LSEs* become 7.0%, 7.52%, 6.36% and 6.36% for QSs 9, 11, 13 and 15, respectively, which are improved only slightly over those in Fig. 9(a) with $M_{el,m}$ = 12. When using $M_{el,m}$ = 20, *LSEs* for these states are reduced significantly from 5.77%, 6.5%, 4.18% and 4.78% in Fig. 9(a) (DoF = 233 in Fig. 9(b)) to 3.56%, 3.02%, 2.75% and 3.15% in Fig. 10(a) (DoF = 250 in Fig. 10(b)), respectively. Even more reduction is observed in Fig. 10(a) when using $M_{el,m}$ = 30 (DoF = 295), where the *LSEs* equal 3.27%, 2.83%, 1.68% and 2.75% for QSs 9, 11, 13 and 15, respectively, and they continue deceasing slowly if a higher dimension $M_p$ for $\mathbf{H}_p$ in (17) is included in and/or large $M_{el,m}$ is used.

While comparing Figs. 9(a) with Fig. 10(a), it is interesting to observe that *LSEs* are nearly identical between these two sets of results in each state for $M_{el,m}$ < 12. Beyond $M_{el,m}$ = 12, use of higher dimension QEM-Hamiltonian matrix in (17) gradually improves the prediction accuracy. This implies that, if the training data quality is not sufficient (e.g., in the extrapolation case), use of more modes may not effectively improve the prediction accuracy unless a larger $M_p$ is also used in solving (17). A larger $M_p$ with a larger $M_{el,m}$ leads to a better accurate prediction of WFs but also slightly increases computational time due to the larger DoF in post processing.

**Structure B: 20 × 20 elements**

One trained generic element, Element B, is used to construct the 20 × 20 elements in Structure B for this demonstration to predict WFs in the first 15 states. Similar to Element A, Element B is represented by one set of POD *U* modes and one set of POD WF modes. Numerical settings for Structure B, training of POD *U* modes for Element B, and variations of $\sigma$ and $(x_o, y_o)$ are described in Secs. 3.1 and 3.2 and Table 1. Notes that Element B includes 3-DoF variation in potential in the training, compared to Element A with only one DoF variation, and both are trained by 50 samples of random potentials, as shown in Table 1. Poorer quality of training WF data and POD WF modes for Element B than Element A is thus expected. For Structure B, 50 samples of WFs in the first 15 states for Element B are collected from DNSs of a domain of 2 × 2 elements (a total of 15 × 50 × 2 × 2 snapshots, same as that for Element A) to generate POD WF modes. In this case, QEM-Galerkin with potential within the training bounds leads to large errors in higher QSs even using a large number of POD WF modes and a higher dimension of $\mathbf{H}_p$ in (17). The poor-quality POD WF modes for Element B due to the incomplete training can certainly be improved if considerably more samples of random potentials are included, for example, by doubling or tripling the random values for each of $\sigma$, $x_o$ and $y_o$ for $U(\vec{r})$ in (25) within the training bounds given in Table 1. This would be computationally intensive in this case with 3-DoF variations in parameters. For the quantum POD-Galerkin methodology, an alternative can be applied to effectively improve the training data quality, which is illustrated below.

According to Item (c) for the training factor in Sec. 3.3, the number of QSs of training WF data is doubled in the demonstration of Structure B to enhance the training data quality; namely, WFs in the first 30 states (only 15 states needed in the prediction) are collected. There are thus 6,000 (30 × 50 × 2 × 2) snapshots of WF training data to generate one set of POD WF modes for Element B. Although more training effort is needed, this actually offers a more accurate prediction with a smaller number of modes in post process (than the training with WFs in only 15 QSs) and minimizes the computing time. An example of the training potential profile in this 20 × 20 structure is shown in Fig. 11(a), and the eigenvalue for Element B is illustrated in Fig. 11(b). It is clearly seen that the Element B eigenvalue declines considerably more slowly than that in Fig. 2(b) for Element A due to more complex potential variation. A slightly higher dimension ($M_p$ = 25) in (17) is thus used ($M_p$ = 20 in Figs. 4-6 for Structure A) is then applied in Structure B to improve the prediction accuracy with a reasonably small number of WF modes.

Periodic BCs are applied to the boundaries of Structure B and the potential profiles are selected with $\sigma$ and $(x_o, y_o)$ randomly assigned within the training bounds given in Table 1. The *LSEs* for the first 15 QSs are illustrated in Fig. 12(a) as functions of $M_{el,m}$, and the needed DoF in post processing for the entire structure is shown in Fig. 12(b). When using $M_{el,m}$ = 15 or 20, it is observed that LSE ≤ 1.55% or 1%, respectively, for all QSs but QSs 1, 4, 6 and 12. For all QSs to reach an LSEs below 2%, $M_{el,m} = 21$ is required and DoF = 340. If LSEs for all QSs below 1.5% is desirable, $M_{el,m} = 23$ and DoF = 358. For the QEM-Galerkin prediction of eigenenergies, errors shown in Fig. 13 for Structure B are higher than those in Fig. 6 for Structure A. However, these errors are still quite small and near or below 0.5% (most of them less than 0.2%) except for QS 2 with an error below 1.1%. The larger errors observed in Structure B, compared to Structure A, are induced not only by more complex potential variation and poorer data quality but also larger potential error resulting from POD $U$ modes with LS fitting for Element B than Element A, as shown in Tables 2 and 3. More POD $U$ modes can be implemented in QEM-Galerkin simulation with more spatial grid points for the LS fitting (only 132 points applied currently) to improve the potential accuracy. Moreover, a higher dimension for $\mathbf{H}_p$ in (17), as presented in Fig. 10 for the extrapolation case, can be also applied to further minimize the LSE.

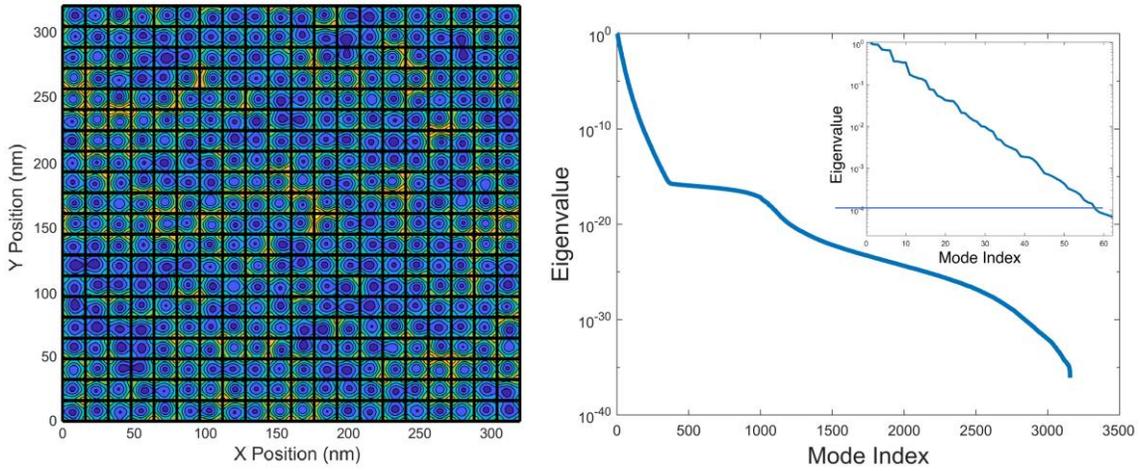

Fig. 11. (a) Contour of a training potential in Structure B with 20 × 20 elements. (b) Eigenvalue in descending order with a zoom-in inset.

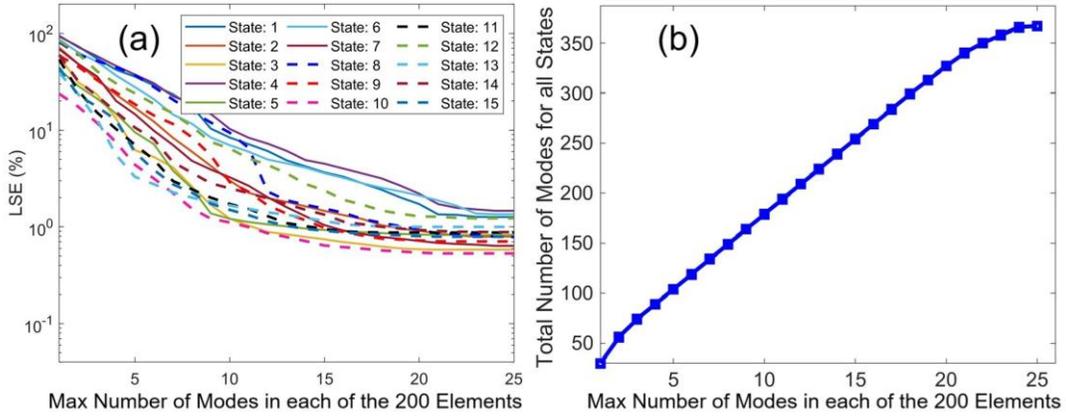

Fig. 12. (a) LSE of QSs 1-15, $N_\mu = 0.01$, as a function of the maximum number of modes ($M_{el,m}$) in each of the 400 elements. (b) Total number of modes (or DoF) used for the first 15 states needed in post processing.

The $|\psi|^2$ profiles along $x$ and $y$ directions, together with the QD potential, through their maximum magnitudes predicted by QEM-Galerkin using $M_{el,m}$ between 10-16 modes, are illustrated in Fig. 14, compared against DNS results. When using a large enough number of modes, the WFs predicted by QEM-Galerkin overlap with DNS results quite well in a very complex 2D potential profile. Zoom-in profiles along *x* and *y* for QS 10 using 1, 5 and 16 modes are given in Fig. 15, where a reasonable prediction is observed with $M_{el,m}$ = 5 (an *LES* = 4.3% given in Fig. 12(a)) and nearly a perfect prediction is achieved with $M_{el,m}$ = 16 and an LES = 0.62%.

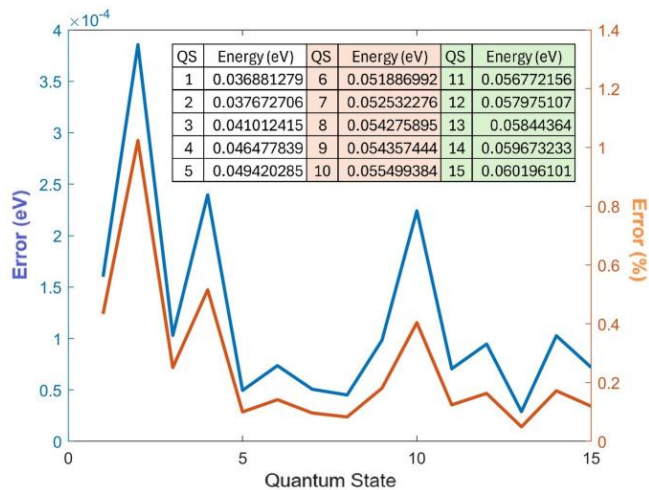

Fig. 13. Errors of the eigenenergies for QSs 1-15 in eV and percentage with respect to the minimum potential energy with $N_\mu$ = 0.01. The inset table shows the energies of the first 15 QSs.

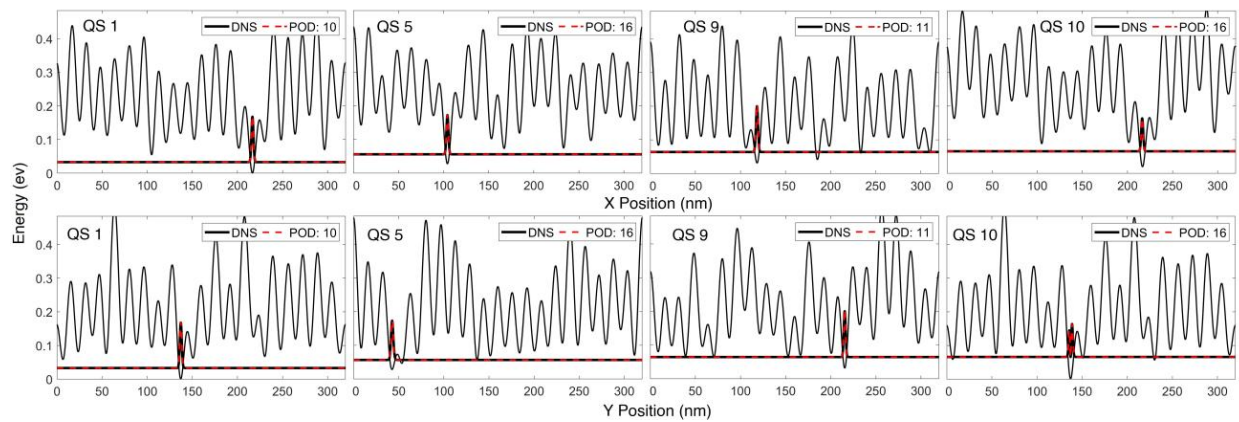

Fig. 14. Potential profiles, together with $|\psi(x,y)|^2$ for QS 1, 5, 9 and 10. The top and bottom rows are along *x* and *y* directions, respectively, through the maximum $|\psi|^2$. The POD mode number in the legend equals $M_{el,m}$.

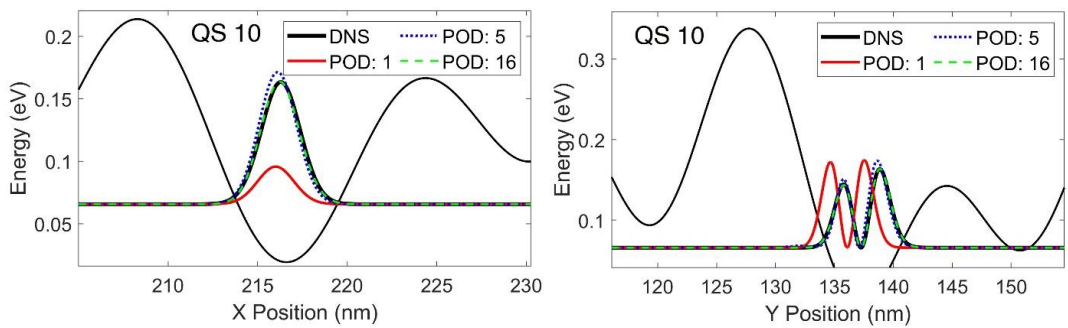

Fig. 15. Zoom-in profiles of QS 10 along *x* and *y* directions.
19

## 4. Discussions

A thorough investigation of QEM-Galerkin for simulations of two QD structures, each of which is constructed by a generic element, has demonstrated the conciseness and resilience of POD bases for representing two distinct natures. These include (i) QD potential profiles whose characteristics are similar to the pseudopotential assumed in DFT calculations, and (ii) electron WFs confined by QDs in two periodic structures formed by $4 \times 4$ and $20 \times 20$ QD elements. To speed up calculations of QD potential energy in POD space, $U_{\eta_p,ij}$, in (14) for QEM-Galerkin simulation, $U(\vec{r})$ is represented by one generic set of POD $U$ modes for all elements in the structure with its weighting coefficients $c_n$ in (21) determined by *LS* fitting. Additionally, one set of POD WF modes for a generic element is trained to represent all elements in the structure, where the IPDG method [34, 35] is applied to impose interface continuity between elements when stitched together to construct the structure. To incorporate first principles, the *ab initio* Galerkin projection of the Schrödinger equation for each test structure is further applied onto the POD space spanned by the generated generic POD WF modes of all the elements.

Important findings and concepts resulting from this investigation of *ab initio* QEM-Galerkin enabled by POD *U* modes and POD WF modes are briefly discussed below.

**POD potential basis:** It has been demonstrated that computational time of the projection of nonuniformly varying potential $U(\vec{r})$ onto WF POD space during QM-Galerkin simulation can be significantly reduced if $U(\vec{r})$ is represented by a set of optimized *U* modes given in (21). Considering both accuracy and efficiency, Tables 2 and 3 in Sec. 3.2 have demonstrated the superiority of the POD *U* modes with *LS* fitting to the Fourier *U* modes using FFT in periodic structures with $4 \times 4$ and $20 \times 20$ QD elements. With similar accuracy, computing the potential in WF POD space using the *LS*-fit *U* mode approach is 15 to 22 times as fast compared to using Fourier *U* modes due to a considerably smaller number of modes needed in the POD *U* basis. For Structures A and B, if an *LSE* for $U(\vec{r})$ near or below 0.12% is desired, the *LS*-fitting with POD *U* modes to calculate $U_{\eta_p,ij}$ is 362 to 386 times more efficient than the direct numerical integrals for $U_{\eta_p,ij}$ in (14). Accuracy using the POD *U* modes can be improved by using a few more modes but for the Fourier basis considerably more modes would be needed even for periodic potentials. Between the two approaches for the POD *U* modes, *LS* fitting in (24) is more than an order faster than the direct projection in (23).

**Generic elements:** As illustrated in this work, a large quantum structure can be modeled accurately by stitching many smaller well-trained elements together, using QEM-Galerkin, with interface continuity enforced by IPDG. This work selects a trained generic element to construct and represent the entire structure. The generic-element practice allows more effective training in a small domain, such as training domains with $2 \times 2$ elements for Elements A and B to represent structures with $4 \times 4$ or $20 \times 20$ elements. In practice, generic elements can be selected as building blocks for a specific category of technology, which can be trained and stored in a library for constructing larger structures. Use of trained generic elements can be applied to improve cost effective design, similar to many engineering and material applications. In addition, effective training of generic elements enables the element QEM-Galkerin model to accept a variety of variations, such as the potential variation in the case of $20 \times 20$ elements. This could offer an effective simulation tool for detailed tolerance tests in nanostructure design to examine unavoidable manufacturing fluctuations. Such tolerance tests based on computationally intensive DNSs are practically prohibitive.

**Extrapolation prediction guided by *ab initio* Galerkin projection:** The POD-Galerkin methodology offers a unique feature to improve the prediction accuracy and extrapolation ability in case training is incomplete and retaining is not feasible. By using a larger Hamiltonian dimension ($M_p$) for each element in (17), the accuracy of QEM-Galerkin WFs increases evidently using the same number of POD WF modes in post processing. This practice enhances the extrapolation capability for QEM-Galerkin, as presented in Fig. 10 ($M_p = 30$) to Fig. 9 ($M_p = 20$) for Structure A outside the training potential variation. Fig. 9 showed that the incomplete training data in this demonstration led to large *LSEs* in WFs in many QSs. The poor training was compensated by using a higher $\mathbf{H}_p$ dimension of $30 \times 30$ for each element to significantly improve the prediction accuracy using the same number of POD WF modes, as illustrated in Fig. 10. It is believed that by incorporating first principles in the model via *ab initio* Galerkin projection in (11), use of the higher dimension Hamiltonian than what is needed in simulation evidently improves the prediction accuracy for the lower QSs even for the case beyond training.

**Training effectiveness enhanced by *ab initio* Galerkin projection:** In the demonstration of Structure B, training potential in Element B did not thoroughly cover the 3-DoF parametric variations. The QEM-Galerkin model generated by the training data of 15 QS WFs for Element B provided an inaccurate prediction even with a large number of POD WF modes and a higher dimension of $\mathbf{H}_p$. The prediction accuracy of the first 15 WFs was significantly improved with a reasonably small number of POD WF modes when these modes are trained using WF data in the first 30 states (instead of 15 states), together with a slightly higher dimension of $\mathbf{H}_p$ in simulation, as illustrated in Figs. 12-15. This is because WF training data in higher QSs embedded in the generated POD WF modes offers additional information on lower-state WFs. The influences of the higher-state information on the lower states were further enhanced and guided by the first principle via the *ab initio* Galerkin projection in simulation to improve the prediction accuracy for lower-state WFs. This offers a very useful feature for QEM-Galerkin when high DoF parametric variations are involved in quantum problems whose data size becomes massively large when more thorough variation in each DoF is needed to collect the training data.

**Computational efficiency:** Computational time for QEM-Galerkin measures 3 separate steps: construction of the $N_{el}$ element QEM-Galerkin Hamiltonian of (17), QEM-Galerkin simulation in POD space solving (17) and post processing via (20) to obtain spatial WFs in the entire structure. These times, compared to DNS, are listed in Table 4 for demonstrations of Structure A (the interpolation case) and Structure B. For Structure A the QEM-Galerkin simulation was performed with $M_{el,m} = 14$ that achieves an *LSE* < 1% for all 15 states, and for Structure B the simulation was carried out with $M_{el,m} = 23$ that leads to an *LSE* < 1.5% for all 15 states. In general, post processing is the most time-consuming calculation since fine resolution was applied to enhance the training quality and the prediction accuracy in this study. Overall, QEM-Galerkin is 198 and 752 times more efficient than DNS for Structures A and B, respectively, and QEM-Galerkin offers more efficiency improvement, compared to DNS, for a structure comprising more elements. If fine resolution is not needed for WFs, coarse resolution WFs can be calculated in (20) to improve the computing efficiency without sacrificing the prediction accuracy.

Table 4 Computational time for QEM-Galerkin, compared to DNS

| Structure | DNS (s) | Hamiltonian construction (s) | QEM-Galerkin simulation (s) | Post process (s) | Improved efficiency |
|---|---|---|---|---|---|
| A: $4 \times 4$ | 31.6 | 0.0542 | 0.0312 | 0.0740 | 198 |
| B: $20 \times 20$ | 2684 | 0.418 | 0.601 | 2.55 | 752 |

## 5. Conclusions

This study has demonstrated that quantum generic elements represented by POD *U* modes and POD WF modes can be used to predict WFs in multiple QSs with high accuracy and efficiency provided the training WF data quality is sufficient. In addition to accuracy and thoroughness of the training data indicated in several studies of POD-based approach [24, 28], this work showed that data quality can be further enhanced by including training WF data in higher QSs than what is actually needed in the QEM-Galerkin prediction. Although this increases the training effort, it could actually make QEM-Galerkin more efficient due to a possible reduction in the number of modes in post processing. When more thorough training is needed to improve data quality by covering an enough combination of high DoF parametric variations, collection of a massive amount of data is needed. In this case, the alternative by collecting training WF data in more states to enhance data quality appears to considerably decrease the training effort. When training is incomplete and further training is not feasible, the prediction accuracy can also be significantly improved by using a higher-dimension Hamiltonian matrix in POD space in QEM-Galerkin simulation. This may also allow a smaller number of POD-WF modes to achieve good accuracy and reduce post processing computing time. These flexibilities to improve accuracy stem from the fact that the simulation models based on QEM-Galerkin are closed by the *ab initio* Galerkin projection. This empowers QEM-Galerkin to be guided by the first principles during simulations.

In addition to POD WF modes for representing the quantum system, a set of POD *U* modes with *LS* fitting is applied to speed up computation for the projection of the real space potential onto POD WF space during simulations. The investigation found the POD *U* modes converge to desired accuracy considerably faster than the Fourier *U* modes even for periodic potentials and result in a speed up around 15 to 22 times over the Fourier *U* modes. Overall, QEM-Galerkin offers an improvement of computational efficiency around 198 and 752 times for nanostructures with $4 \times 4$ and $20 \times 20$ elements, respectively, compared to DNS. Also, efficiency is improved considerably more in nanostructures constructed by more elements. In addition, for nanostructures comprising a large number of elements, the QEM-Hamiltonian equation in (17) becomes highly sparse, which will be greatly benefited by parallel computing. With such an efficient and accurate approach, QEM-Galerkin can potentially be applied to develop effective simulation models for electronic and photonic nanostructure and materials. These, for example, include DFT simulations of supercells (similar to the QD structure in Fig. 11), efficient and accurate tolerance analysis for nanostructure performance influenced by inevitable manufacturing fluctuations (such as the potential width and shift variations in Fig. 11), photonic crystals and metamaterials (whose eigenvalue systems and solutions are described by the Helmholtz equation that resembles the Schrödinger equation), etc.

**Acknowledgements:**
This work was supported by National Science Foundation (NSF) under Grant number OAC-2118079 and Clarkson Ignite Graduate Research Fellowships.